\documentclass[conference]{IEEEtran}
\IEEEoverridecommandlockouts
\usepackage{cite}
\usepackage{amsmath,amssymb,amsfonts}
\usepackage{algorithmic}
\usepackage{graphicx}
\usepackage{textcomp}
\usepackage{xcolor}
\usepackage{cite}
\usepackage{amsmath,amssymb,amsfonts}
\usepackage{algorithmic}
\usepackage{graphicx}
\usepackage{tabularray}
\usepackage{textcomp}
\usepackage{xcolor}
\usepackage{geometry}
\usepackage{float}
\usepackage{xcolor}
\usepackage{colortbl,booktabs}
\usepackage{subfig}
\usepackage{subcaption}
\usepackage{float}
\usepackage{multirow}
\usepackage{color}
\usepackage{lipsum}
\usepackage{mathtools}
\usepackage{cuted}
\usepackage{nicematrix}
\usepackage{textcomp}
\usepackage{bm}
\usepackage{svg}
\usepackage{algorithm2e}
\usepackage{dblfloatfix}
\usepackage{siunitx}
\RestyleAlgo{ruled}
\def\BibTeX{{\rm B\kern-.05em{\sc i\kern-.025em b}\kern-.08em
    T\kern-.1667em\lower.7ex\hbox{E}\kern-.125emX}}
\begin{document}

\title{Operational Resilience Assessment: A Frequency-Domain Approach for DC Microgrids \\
 }

\author{\IEEEauthorblockN{\textsuperscript{} Ali Hosseinipour, Maral Shadaei, Javad Khazaei}
\IEEEauthorblockA{\textit{Department of Electrical and Computer Engineering} \\
\textit{Lehigh University}\\
Bethlehem PA, USA \\
Emails: \textit{alh621@lehigh.edu, masb22@lehigh.edu, khazaei@lehigh.edu}}

\thanks{This research was in part under support from the Department of Defense, Office of Naval Research award number N00014-23-1-2402.}
}

\maketitle

\begin{abstract}
DC shipboard microgrids (SMGs) are highly dynamic systems susceptible to failure due to various cyber-physical disturbances, such as extreme weather and mission operations during wartime. In this paper, the real-time operational resilience (OR) evaluation of DC SMGs against dynamic disturbances is proposed using frequency-domain metrics. To this end, first the drawbacks of time-domain OR evaluation using an energy imbalance index is discussed. As the time-domain energy imbalance index is shown to be incapable of real-time OR assessment, particularly in the context of droop-controlled DC SMGs without secondary voltage restoration control, the $\mathcal{H}_2$ and $\mathcal{H_\infty}$ norms of candidate transfer functions (TFs) of the system are proposed as measures of resilience. It is shown that the proposed norms calculated for the bus impedance TF of the system provides intuitive results in terms of energy imbalance and can be computed in real time. The case studies conducted for the study DC SMG under pulsed power load (PPL) disturbances demonstrate the shortcoming of the time-domain OR evaluation and the capability of the proposed frequency-domain metrics in intuitive OR evaluation of DC SMGs.
\end{abstract}


\section{Introduction}\label{sec.introduction}
Shipboard microgrids (SMGs) have been traditionally based on AC architecture due to its mature technology. However, inadequacies of the AC architecture such as the fixed-speed operation of generators, unwanted reactive power flow, imbalances, harmonics, and the bulky nature of transformers reduce the efficiency of power conversion in these systems~\cite{dcship}. These challenges have motivated the application of DC distribution on SMGs due to its flexible architecture, elimination of reactive voltage drop, reduction of weight, and obviation of the phase angle synchronization~\cite{mvdc}. 

Resilience of SMGs is of high importance because of the need to sustain operations under abnormal cyber-physical events such as adversary attacks and electrical faults~\cite{cost_trade}. Resilience is deemed as one of the pillars of energy security by naval facilities engineering systems command (NAVFAC) and is defined as ``the ability to prepare for and recover
from energy disruptions to ensure energy availability and reliability for mission assurance and readiness"~\cite{navfac}. As such, multiple studies have investigated resilience quantification for SMGs. Various topologies of DC SMGs are compared in terms of resilience against worst-case attacks by metrics such as the minimum number of attacks that render the system out of service and the amount of load shed during the attacks~\cite{topology}. Maximization of graph algebraic connectivity and weighted sum of the maximum flow from generator
buses to load buses are proposed as resilience indices to develop line expansion strategies for DC SMGs in~\cite{graph}. In~\cite{mission}, a resilience metric is defined by the expected electrical disruption mission impact to quantify the impact of disruptions in a military microgrid on the base mission, which leverages the classification of loads into critical and non-critical categories in relation to the mission. A resilience metric based on the system invulnerability and recoverability is established in~\cite{cost_trade} and used together with a cost model for the architectural design of a military microgrid. Various resilience metrics defined for DC SMGs are ranked in~\cite{corr} to identify attributes that are the best indicators of the system performance subject to contingencies.

The above-mentioned methods mostly focus on high-impact low-probability events for resilience quantification. As such, these methods are incapable of being assessed in real time since they depend on simulating a multitude of disruptive scenarios to calculate the resilience indices. The notion of operational resilience (OR) is introduced to quantify the resilience against the uncertainty in the generation and load demand in addition to high-impact disruptive events. An adaptive real-power capacity index is proposed in~\cite{or_smr} by aggregating the flexibility curves of dispatchable assets in a hybrid microgrid to evaluate OR in real time. The adaptive real-power capacity provides insight into the size of disturbance the system can tolerate. Similar indices based on adaptive power capacity are proposed in~\cite{or_pvbess} to evaluate the resilience contribution of photovoltaic (PV) and battery energy storage (BESS) power plants and in~\cite{or_distribution} to evaluate OR for power distribution networks. However, the OR evaluation methods in~\cite{or_smr,or_pvbess,or_distribution} fail to properly account for quantifying the short-term contribution of assets to dynamic disturbances. To address this, the moment of inertia is taken into account to derive flexibility curves of generation assets in~\cite{or_inertia}. A resilience monitoring framework employing metric temporal logic is also developed for microgrids in~\cite{formal} to assess the vulnerability of various nodes of the system in the face of cyber-physical anomalies. In spite of the ability of the index in~\cite{formal} to be evaluated in real time, its dependence on the size of disturbance can result in a false sense of resilience level. 

In this paper, a resilience metric is proposed for isolated-low-inertia DC SMGs to assess the system vulnerability towards highly dynamic disturbances such as pulsed-power loads (PPLs) that lead to short-term power imbalance. The main contributions of the paper can be summarized as
\begin{itemize}
    \item Using the concept of energy imbalance, time-domain and frequency-domain metrics are proposed to assess the resilience of DC SMGs against dynamic disturbances.
    \item The drawback of the time-domain resilience metric is demonstrated in the absence of the secondary control in droop-controlled DC SMGs.
    \item Frequency-domain-based metrics, namely $\mathcal{H}_\infty$ and $\mathcal{H}_2$ norms, are proposed and compared for system candidate transfer functions (TFs) for a sensible evaluation of OR level.
    \item The $\mathcal{H}_\infty$ and $\mathcal{H}_2$ norms of the DC SMG bus impedance is proposed as the solution for the real-time assessment of OR in DC SMGs because of being intuitive in terms of energy imbalance, high sensitivity to resilience level, the ability to be computed in real time owing to its data-driven nature, and being independent of disturbance size.
    \item An impedance-based framework is devised for real-time OR evaluation and is tested under various virtual inertia levels of the DC SMG during PPL disturbances.
\end{itemize}
The remainder of the paper is organized as follows. The dynamic modeling of the study DC SMG is addressed in Section~\ref{sec.modeling}. Section~\ref{sec.or} introduces the concept of energy imbalance for OR evaluation and its relation with the system virtual inertia level. The real-time frequency-domain-based OR evaluation framework is proposed in Section~\ref{sec.framework}. Section~\ref{sec.simulation} presents simulation results for the validation of the proposed OR evaluation framework. Finally, Section~\ref{sec.conclusion} concludes the paper.
\section{Time-domain Dynamic Modeling of the DC SMG}\label{sec.modeling}
\begin{figure}
\centering 
\includegraphics[scale=0.75]{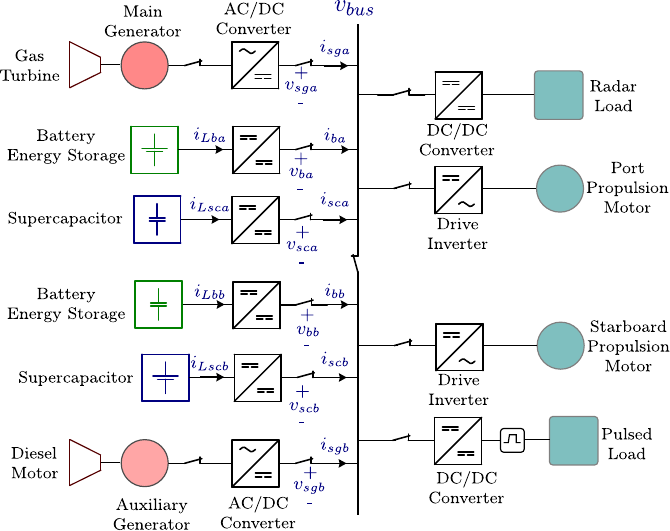}
\caption{Notional DC SMG}
\label{fig.smg}
\end{figure}

\begin{figure}
\centering 
\includegraphics[scale=0.60]{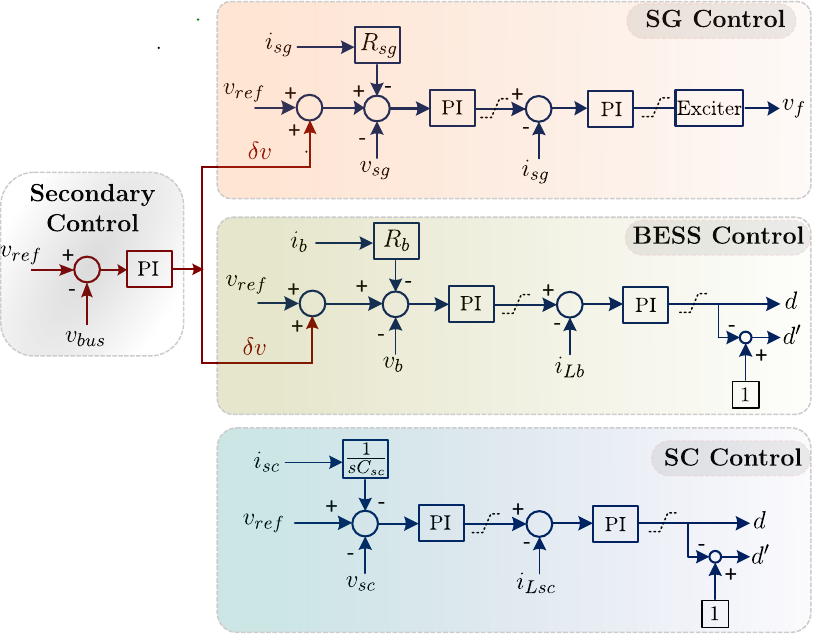}
\caption{Primary controllers of the DC SMG}
\label{fig.controllers}
\end{figure}
The notional diagram of a DC SMG is demonstrated in Fig.~\ref{fig.smg}, where one of the synchronous generators (SGs) supply the majority of the load and the other one act as  the auxiliary unit. Two BESS units are deployed to respond to fast-varying loads, while SCs serve as short-term energy storage to reduce the current stress on BESSs and assist in supplying PPLs. All the power sources in the DC SMG are operating under voltage mode control with cascaded controllers as demonstrated in Fig.~\ref{fig.controllers}. The control input for the BESS power converters are the duty cycles of their half-bridge converters, namely $d$ and $d'$. The control input for the SGs is the field voltage, denoted by $v_f$. Neglecting the fast dynamics of inner voltage and current controllers in Fig.~\ref{fig.controllers}, the output voltage of SG and BESS units is decided by
\begin{align}\label{eq.droopp}
  v_k=v_\text{ref}-R_{k}i_{k}, ~\forall k \in\{\text{sga, sgb, ba, bb}\}
\end{align}
where $v_{\text{ref}}$ is the voltage setpoint, output currents of the SGs and BESSs are denoted by 
 $i_k$, and $R_k$ represents the droop gain for SGs and BESSs.
 
In order for the SCs to respond to transient power fluctuations, an integral droop control~\cite{integral_droop} decides their voltage reference as
\begin{align}\label{eq.droop}
  v_h=v_\text{ref}-\frac{1}{C_{h}}\int i_{h}, ~\forall h \in\{\text{sca, scb}\}
\end{align}
where $C_h$ is the virtual capacitance of the SCs, and $i_h$ denotes the output current of the SCs.
 
\begin{figure}
\centering 
\includegraphics[scale=0.80]{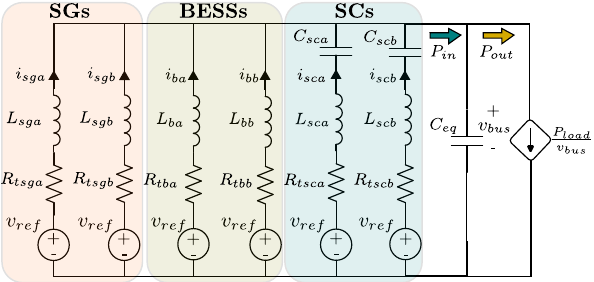}
\caption{Reduced-order model of the DC SMG}
\label{fig.simplified_multi}
\end{figure}

The reduced-order model of the DC SMG of Fig.~\ref{fig.smg} can then be represented by Fig.~\ref{fig.simplified_multi}~\cite{multiconverter}. Owing to the high-bandwidth control of point-of-load converters in Fig.~\ref{fig.smg}, they can be represented by a lumped constant power load represented by the dependent current source in Fig.~\ref{fig.simplified_multi}. The system of differential equations governing the dynamic behavior of the DC SMG are resulted below after writing the node and loop equation for the circuit of Fig.~\ref{fig.simplified_multi}.
\begin{align}{}
C_{\text{eq}} \frac{dv_\text{bus}}{dt} &= \sum\limits_{m\in \mathcal{N}}i_{{m}} - \frac{P_{\text{load}}}{v_\text{bus}}, ~\mathcal{N} =\{\text{sga, sgb, ba, bb, sca, scb}\}
\label{eqn.capp_voltage} 
\\
    L_k\frac{di_{k}}{dt} &= v_{\text{ref}} - R_{tk}i_{k} - v_\text{bus}, ~\forall k \in\{\text{sga, sgb, ba, bb}\}
  \label{eq.sga_current} 
  \\
    L_{h}\frac{di_{h}}{dt} &= v_{\text{ref}} - r_hi_{h} - v_{z} - v_\text{bus}\\
C_{h}\frac{dv_{z}}{dt} &= i_{h}, ~\forall h \in\{\text{sca, scb}\},~\forall z \in\{\text{$C_\text{sca}$, $C_\text{scb}$}\}
\end{align}
where $v_\text{bus}$ is the voltage across the DC link capacitor $C_{\text{eq}}$. The equivalent inductance of the SG and BESS branches is denoted by $L_k$. The equivalent inductance of the SC branches is denoted by $L_h$. The virtual capacitance of the SCs is represented by $C_h$. The voltage across the virtual capacitors is denoted by $v_z$. The branch resistance of SGs and BESSs is given as $R_{tk}=R_k+r_k$, where $r_k$ is the aggregate line and parasitic resistance of the branch. The aggregate line and parasitic resistance for SC branches is denoted by $r_h$. The aggregate load power is given by $P_\text{load}$.

Equation (\ref{eqn.cap_voltage}) makes the system of equations nonlinear. Applying the Jacobian linearization technique, the linear equations are obtained as 
\begin{align}{}
C_{\text{eq}} \frac{d\Delta v_\text{bus}}{dt} &= \sum\limits_{m\in \mathcal{N}}\Delta i_{{m}} + \frac{P_{\text{load}}^*}{V_\text{bus}^{*^2}},\\\nonumber\mathcal{N} &=\{\text{sga, sgb, ba, bb, sca, scb}\}
\\\label{eqn.cap_voltage} 
    L_k\frac{d\Delta i_{k}}{dt} &=- R_{tk}\Delta i_{k} - \Delta v_\text{bus}, ~\forall k \in\{\text{sga, sgb, ba, bb}\} 
  \\\label{eq.sga_current}
    L_{h}\frac{d\Delta i_{h}}{dt} &=- r_h\Delta i_{h} - \Delta v_{z} - \Delta v_\text{bus}\\
C_{h}\frac{d\Delta v_{z}}{dt} &= \Delta i_{h}, ~\forall h \in\{\text{sca, scb}\},~\forall z \in\{\text{$C_\text{sca}$, $C_\text{scb}$}\}
\end{align}
where $P^*_{load}$ and $V^*_\text{bus}$ denote the equilibrium load power and DC bus voltage, respectively.
\section{Inertia Equivalence and OR in DC SMGs}\label{sec.or}
The DC link power balance equation in the DC SMG of Fig.~\ref{fig.simplified_multi} can be written as
\begin{align}\label{eq.dclink}
P_\text{in}-P_\text{load}=C_\text{eq} v_\text{bus}\frac{dv_\text{bus}}{dt}\approx C_\text{eq} v_\text{n}\frac{dv_\text{bus}}{dt}
\end{align}
where $P_\text{in}$ and $P_\text{load}$ are the powers flowing in and out of the DC link capacitor branch as depicted in Fig.~\ref{fig.simplified_multi}. The nominal DC link voltage is signified by $v_\text{n}$.

An analogy can be made with AC power systems where the swing equation governs the frequency dynamics by
\begin{align}  \label{eq.swing}
P_\text{m}-P_\text{e}-D(\omega-\omega_\text{n})=J\omega\frac{d\omega}{dt} \approx J\omega_\text{n}\frac{d\omega}{dt}
\end{align}
where $J$ is the total moment of inertia, $\omega$ is the grid angular frequency, $P_\text{m}$ is the mechanical power, $P_\text{e}$ is the electrical power, $D$ is the damping factor, and $\omega_\text{n}$ is the nominal angular frequency of the grid.

Comparing (\ref{eq.dclink}) and (\ref{eq.swing}), it can be concluded that the DC link capacitance $C_\text{eq}$ represents the virtual inertia in the DC SMG~\cite{dcinertia}. Further analysis of (\ref{eq.dclink}) can provide insight into how load disturbances affect the transient response of $v_\text{bus}$. Assuming a positive step disturbance in $P_\text{load}$ at time t=0 as the result of PPL change in the DC SMG, the voltage nadir can be defined by
\begin{align}
    |\underbar{$v_\text{bus}$}|\triangleq \min_{t\geq0} v_\text{bus}(t)
\end{align}

Furthermore, the rate of change of voltage (RoCoV) can be defined by
\begin{align}
    |\dot{v}_\text{bus}|_\text{max}\triangleq \max_{t\geq0} |\dot{v}_\text{bus}(t)|
\end{align}

From (\ref{eq.dclink}), it can be concluded that 
\begin{align}\label{eq.prop}
 |\underbar{$v_\text{bus}$}|\propto |\dot{v}_\text{bus}|_\text{max}\propto |\Delta P_\text{load}| \propto \frac{1}{C_\text{eq}}  
\end{align}

Equation (\ref{eq.prop}) verifies the credibility of $v_\text{bus}$ as a performance output to be utilized in an energy imbalance metric that can represent the OR for the DC SMG. The energy imbalance can thus be computed during a disturbance by~\cite{inertia_placement}
\begin{align}\label{eq.Ev}
    E_{v}=\int_{\tau_1}^{\tau_2} {\Delta\bm{v}_\text{bus}}^T\Delta\bm{v}_\text{bus} dt
\end{align}
where $\Delta\bm{v}_\text{bus}=[\Delta v_{\text{bus},1}, \Delta v_{\text{bus},2}, ...]^T$ is the time-domain evolution of $\Delta v_\text{bus}(t)$ over the time-window $\tau_1<t<\tau_2$, where $\tau_1$ is a time before the disturbance and $\tau_2$ is a time after the disturbance.

Using $E_v$ as a metric for the resilience level is not without challenges. The first challenge is that $E_v$ is disturbance-dependent. This means that a larger 
$\Delta P_\text{load}$ leads to a higher OR index while the system characteristics relevant to the resilience such as $C_\text{eq}$ remain unchanged. The second challenge is that in droop-controlled DC SMGs, the post-disturbance steady-state voltage deviation will dominate $\Delta v_\text{bus}$ and leads to misleading results for energy imbalance. The real-time OR assessment using $E_v$ is also an impossible task since not only it is dependent on the occurrence of disturbances in the system but also due to the temporal nature of $E_v$, enabling only a quasi-real time computation of $E_v$.

\section{Real-time Frequency-domain OR Assessment}\label{sec.framework}
As explained in the previous section, the energy imbalance metric may lead to misleading results regarding the OR level and also does not have the capability to be evaluated in real time. In this section, the $\mathcal{H}_2$ norm is proposed to evaluate the OR level for DC SMGs. The $\mathcal{H}_2$ norm is interpreted as the energy of the system response to an impulse disturbance. Since the impulse response is the time derivative of the step response, the $\mathcal{H}_2$ norm is dominated by the transient response of the system as the step response derivative of the performance output has a finite limit as $t\rightarrow\infty$~\cite{inertia_placement}. Since the calculation of the $\mathcal{H}_2$ norm in the time domain requires the knowledge of the system dynamics, we propose the $\mathcal{H}_2$ calculation in frequency-domain using TFs of the system. The advantage of using TFs is that they are independent of specific disturbances and can be computed in real time by frequency response data collection.

Two candidate TFs are proposed and compared for OR evaluation in this study, namely the load-power-to-output-voltage TF, $G_\text{pv}(s)$, and the bus impedance TF, $Z_\text{bus}(s)$. The former TF can be analytically obtained as
\begin{align}\label{eq.Gpv}
        G_\text{pv}(s)&=\frac{\Delta v_\text{bus}}{\Delta P_\text{load}}
        =\frac{C_\text{eq}}{P^*_\text{load}+{V^*_\text{o}}^2(Y_\text{t}(s)-C_\text{eq}s)}
\end{align}
where
\begin{align}     Y_\text{t}(s)&=\sum\limits_{n\in \mathcal{N}}Y_{{n}}, ~\mathcal{N} =\{\text{sga, sgb, ba, bb, sca, scb}\}\\
        Y_{k}&=\frac{-1}{L_{k}s+R_{k}}, ~\forall k \in\{\text{sga, sgb, ba, bb}\}\\
        Y_{h}&=\frac{-C_{h}s}{L_{h}C_{h}s^2+R_{h}C_{h}s+1}, ~\forall h \in\{\text{sca, scb}\}
\end{align}

The bus impedance transfer function can also be obtained as the parallel equivalent impedance of all branches in Fig.~\ref{fig.simplified_multi} by
\begin{align}     Z_\text{bus}(s)&=\frac{1}{\sum\limits_{n\in \mathcal{N}}|Y_{{n}}|+C_{eq}s-\frac{P^*_\text{load}}{V^*_\text{o}}}, \\\nonumber~\mathcal{N} &=\{\text{sga, sgb, ba, bb, sca, scb}\}
\end{align}

The $\mathcal{H}_2$ norm of a stable continuous system with transfer function $G(s)$ is the root-mean square of its impulse response, obtained by
\begin{align}\label{eq.h2}
\|G\|_{\mathcal{H}_2}&=\sqrt{\frac{1}{2 \pi}\int_{-\infty}^{\infty} \operatorname{Tr}(G(j\omega)^* G(j\omega)) \,d\omega }
\end{align}
\begin{figure}
    \centering
    \includegraphics[width=0.95\linewidth]{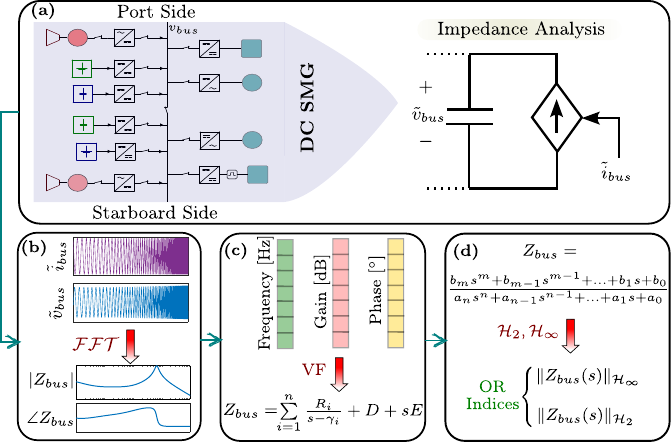}
    \caption{Frequency-domain-based OR evaluation framework for DC SMGs.}
    \label{fig.framework}
\end{figure}

Although not as intuitive as ${\mathcal{H}_2}$ norm in terms of energy imbalance, the ${\mathcal{H}_\infty}$ norm of the DC SMG TFs is also investigated as a potential index for real-time OR evaluation. The ${\mathcal{H}_\infty}$ norm of a stable dynamical system with transfer function $G(s)$ returns
\begin{align}\label{eq.hinf}
\|G\|_{\mathcal{H}_\infty}&=\max_{\omega \in \mathcal{R}}|G(j\omega)|
\end{align}
Due to the distributed nature of the loads in the microgrid, real-time parametrization of (\ref{eq.Gpv}) is challenging. However, real-time measurement of $Z_\text{bus}$ can be realized by online impedance analysis as shown in Fig.~\ref{fig.framework}. To this end, the DC bus of the SMG is perturbed with a signal generator of Fig.~\ref{fig.framework}(a) excited by a swept sine wave as demonstrated in Fig.~\ref{fig.framework}(b). The frequency response data including the perturbation frequency, gain, and phase are then obtained by fast Fourier transform (FFT) and fed to a vector fitting (VF) algorithm~\cite{vf} as depicted in Fig.~\ref{fig.framework}(c). The VF algorithm fits a TF to the frequency response data by a rational function in the form of
\begin{align}
Z_{\text{bus}}(s) &=\frac{\tilde{v}_\text{bus}(s)}{\tilde{i}_\text{bus}(s)}= \frac{b_m s^m + b_{m-1}s^{m-1} + \ldots + b_1s + b_0}{a_n s^n + a_{n-1}s^{n-1} + \ldots + a_1s + a_0} \\\nonumber
&= \sum_{i=1}^n \frac{R_i}{s-\lambda_i} + D + sE
\end{align}
where $R_i$ is the residue corresponding to eigenvalue $\lambda_i$.
\section{Case studies}\label{sec.simulation}
The reduced-order model of the DC SMG presented in Fig.~\ref{fig.simplified_multi} is simulated in Simulink\textsuperscript{\textregistered} with the parameters given in Table~\ref{tab.param}. Then by simulating PPL variations, first the shortcoming of the time-domain index $E_v$ in OR assessment is discussed and then the $\mathcal{H}_2$ and $\mathcal{H}_\infty$ norms of candidate TFs are compared and the $Z_\text{bus}$ is shown as the ideal TF for real-time OR assessment purposes.

\begin{table}
\renewcommand{\arraystretch}{1.3}
\caption{\footnotesize {Simulation parameters of the DC SMG.}}
\label{tab.param}
\centering
\begin{tabular}{c c}
\hline\hline
\bfseries {Parameter} & \bfseries {Value}\\
\hline
$L_{sga},~L_{sgb}$ & \SI{1}{\milli\henry}\\
$L_{ba},~L_{bb}$ & \SI{0.8}{\milli\henry}\\
$L_{sca},~L_{scb}$ & \SI{0.4}{\milli\henry}\\
$R_{sga},~R_{sgb}$ & \SI{0.05}{\ohm}, \SI{0.1}{\ohm}\\
$R_{ba},~R_{bb}$ & \SI{0.225}{\ohm}, \SI{0.45}{\ohm}\\
$r_{sga},~r_{sgb}$,~$r_{ba},~r_{bb},~r_{sca},~r_{scb}$ & \SI{0.05}{\ohm}\\
$C_{sca},~C_{scb}$ & \SI{5}{\farad}, \SI{10}{\farad}\\
$C_{eq}$ & \SI{100}{\milli\farad}\\
$v_{n}$ & \SI{6}{\kilo\volt}\\
\hline\hline
\end{tabular}
\end{table}

\subsection{Performance of $E_v$ for time-domain OR evaluation}
\begin{figure}
    \centering
    \includegraphics[width=1\linewidth]{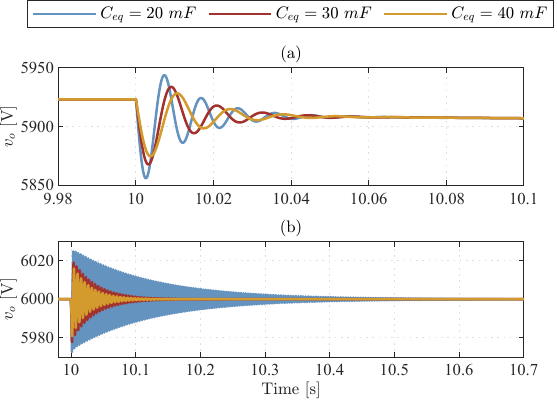}
    \caption{Impact of $C_\text{eq}$ on voltage transient behavior of DC SMGs (a) w/o secondary control (b) w/ secondary control.}
    \label{fig.Ceffect}
\end{figure}

In this section, $E_v$ is calculated for three virtual inertia levels of the DC SMG for two different scenarios. One without secondary control, and the other one with a secondary control that restores the nominal voltage of the DC SMG. It is worth noting that $E_v$ is calculated using real-time $v_{bus}$ measurements instead of $\Delta v_{bus}$, which can be challenging to obtain. Fig.~\ref{fig.Ceffect}(a) and Fig.~\ref{fig.Ceffect}(b) show the voltage transient response after a step-up disturbance of 5~MW in $P_\text{load}$ at $t=$\SI{10}{s}. The calculation of $E_v$ for the time-windows shown in these figures leads to the values presented in Table~\ref{tab.Ev}. In the case w/o secondary control, the results are counter-intuitive since $E_v$ increases with increased $C_{eq}$, implying higher energy imbalance in the DC SMG. This error stems from the post-disturbance voltage deviation caused by droop controllers in this case. However, in the case w/ secondary control, the trend of $E_v$ is decreasing as $C_{eq}$ increases, indicating a drop in energy imbalance, which is consistent with the transient simulation results in Fig.~\ref{fig.Ceffect}.
\begin{table*}
\renewcommand{\arraystretch}{1.3}
\caption{\footnotesize {{Calculation of $E_v$ based on $v_\text{bus}$ measurements for different values of $C_\text{eq}$ w/ and w/o secondary control.}}}
\label{tab.Ev}
\centering
\begin{tabular}{c c c c c c c c c}
\hline\hline
          & \multicolumn{3}{c}{W/O Secondary Control}     &  & \multicolumn{3}{c}{W/ Secondary Control}          \\ \cline{2-4} \cline{6-8} 
          & $C_{eq}$=20 mF & $C_{eq}$=30 mF & $C_{eq}$=40 mF &  & $C_{eq}$=20 mF & $C_{eq}$=30 mF & $C_{eq}$=40 mF \\ \hline
$E_{v}$  &  $7.10295 \times 10^{11}$              &   $7.10297 \times 10^{11}$             &   $7.10299 \times 10^{11}$             &  & $7.3439538 \times 10^{11}$               &     $7.3439504 \times 10^{11}$           &     $7.3439500 \times 10^{11}$           \\ \hline\hline
\end{tabular}
\end{table*}

\subsection{Frequency-domain OR Evaluation}
\begin{figure}
    \centering
    \includegraphics[width=1\linewidth]{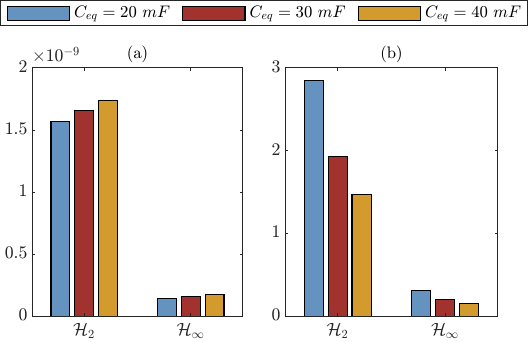}
    \caption{$\mathcal{H}_2$ and $\mathcal{H}_\infty$ norms of (a) $G_\text{pv}(s)$, (b) $Z_\text{bus}(s)$ at different $C_\text{eq}$ values.}
    \label{fig.H}
\end{figure}

\begin{figure}
    \centering
    \includegraphics[width=1\linewidth]{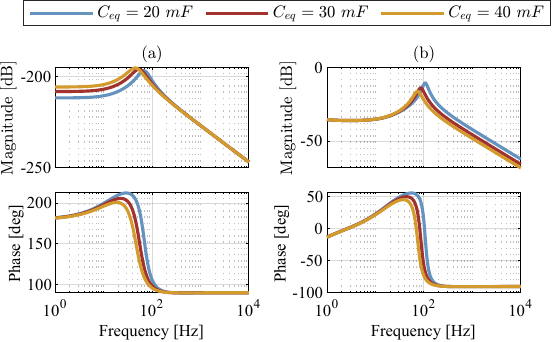}
    \caption{Bode plots of (a) $G_\text{pv}(s)$, (b) $Z_\text{bus}(s)$ at different $C_\text{eq}$ values.}
    \label{fig.bode}
\end{figure}

The frequency-domain OR assessment using the indices proposed in Section~\ref{sec.framework} is discussed in this section for the study DC SMG without the secondary control.  The $\mathcal{H}_2$ and $\mathcal{H}_\infty$ norms of $G_\text{pv}(s)$ and $Z_\text{bus}(s)$ are given in Fig.~\ref{fig.H}. It can be seen that the variation in the norms calculated for $G_\text{pv}(s)$ subject to different virtual inertia levels is negligible. Moreover, the $\mathcal{H}_2$ norm for $G_\text{pv}(s)$ is increasing with increased $C_\text{eq}$, which is counter-intuitive in terms of energy imbalance. However, the $\mathcal{H}_2$ norm of $Z_\text{bus}(s)$ shows continuous decrease at higher values of $C_\text{eq}$, implying reduced energy imbalance, which is consistent with increased virtual inertia level of the system. The decrease in the $\mathcal{H}_\infty$ norm of $Z_\text{bus}(s)$ is also significant, which is also verified in Fig.~\ref{fig.bode}(b) as the maximum value of the bus impedance magnitude is attenuated at higher values of $C_\text{eq}$. These in addition to the capability of $Z_\text{bus}(s)$ to be measured in real time, makes  $\|Z_\text{bus}\|_2$ and $\|Z_\text{bus}\|_\infty$ ideal candidates for real-time OR assessment of DC SMGs.





\section{Conclusion}\label{sec.conclusion}
The present study investigates the real-time OR assessment for DC SMGs subject to physical dynamic disturbances such as PPLs. It is shown that the energy imbalance index evaluated in time domain provides valid OR assessment in real time only for DC SMGs with secondary control and fails in the case without secondary control as the post-disturbance deviation caused by droop control leads to an error in the energy imbalance index. Therefore, the $\mathcal{H}_2$ and $\mathcal{H}_\infty$ norms of the bus impedance TF is proposed for OR assessment, which result in indices that are disturbance independent, and more importantly can be evaluated in real time using impedance analyzers. These indices can thus be employed for developing optimization and control algorithms to enhance real-time OR level in DC SMGs. 

\bibliographystyle{IEEEtran}
\bibliography{IEEEabrv,conference_101719}
\vfill

\end{document}